\documentclass[aip,reprint]{revtex4-1}

\usepackage{graphicx}
\usepackage{epstopdf}
\usepackage{pslatex}
\usepackage{dcolumn}
\usepackage{bm}
\usepackage{color}
\usepackage[dvipsnames]{xcolor}
\usepackage{siunitx}
\usepackage{soul}

\begin{document}

\title{A versatile setup for studying size and charge-state selected polyanionic nanoparticles}
  
\author{K. Raspe}
\affiliation{Institute of Physics, University of Rostock, 18059 Rostock, Germany}

\author{M. M\"uller}
\affiliation{Institute of Physics, University of Rostock, 18059 Rostock, Germany}
\affiliation{Institute of Physics, University of Greifswald, 17489 Greifswald, Germany}

\author{N. Iwe}
\affiliation{Institute of Physics, University of Rostock, 18059 Rostock, Germany}
  
\author{R. N. Wolf}
\affiliation{Institute of Physics, University of Greifswald, 17489 Greifswald, Germany}
\affiliation{\emph{present adress} 
ARC Centre for Engineered Quantum Systems, University of Sydney, Australia}

\author{P. Oel\ss ner}
\affiliation{Institute of Physics, University of Rostock, 18059 Rostock, Germany}

\author{F. Martinez}
\affiliation{Institute of Physics, University of Rostock, 18059 Rostock, Germany}

\author{L. Schweikhard}
\affiliation{Institute of Physics, University of Greifswald, 17489 Greifswald, Germany}

 \author{K.-H. Meiwes-Broer}
\affiliation{Institute of Physics, University of Rostock, 18059 Rostock, Germany} 
\affiliation{Department of Life, Light and Matter, University of Rostock, 18059 Rostock, Germany }

\author{J. Tiggesb\"aumker}
\email{josef.tiggesbaeumker@uni-rostock.de}
\affiliation{Institute of Physics, University of Rostock, 18059 Rostock, Germany}
\affiliation{Department of Life, Light and Matter, University of Rostock, 18059 Rostock, Germany }

\begin{abstract}

Using the example of metal clusters, an experimental setup and procedure is presented, which allows for the generation of size and charge-state selected polyanions from monoanions in a molecular beam. As a characteristic feature of this modular setup, the further charging process via sequential electron attachment within a 3-state digital trap takes place after mass-selection. In contrast to other approaches, the rf based concept permits to access heavy particles. The procedure is highly flexible with respect to the preparation process and potentially suitable for a wide variety of anionic species. By adjusting the storage conditions, i.e., the radio frequency, to the change in the  mass-to-charge ratio, we succeeded to produce clusters in highly negative charge states, i.e., Ag$_{800}^{7-}$.  The capabilities of the setup are demonstrated by experiments extracting electronic and optical properties of polyanionic metal clusters by analyzing the corresponding photoelectron spectra. 

\end{abstract}

\maketitle

\section{Introduction}

An atom in the gas phase cannot bind more than a single extra electron as screening does not compensate for the resulting Coulomb repulsion. This changes, when larger entities are considered, whereby the extra electrons cause the formation of a Coulomb barrier.\cite{SchS95} This leads to an interesting scenario, in which some of the electrons may occupy levels above the vacuum energy, i.e. metastable states characterized by a negative electron affinity.\cite{WanN99} Examples of polyanions are organic molecules\cite{WanARPC09} and clusters of simple metals\cite{YanPRL01} representing very different systems, i.e., the extra electrons are located at specific sites or are completely delocalized within the particle mean-field potential, respectively. The properties and dynamics of polyanionic molecules were studied extensively,\cite{WanJCP15,HorJPCL12,VonPCCP14} with one of the main motivations arising from physical chemistry, i.e., ion solvation in liquids.\cite{StaJPCA02}

Metal cluster polyanions have been investigated in detail in terms of the formation\cite{HerPS99} and fragmentation.\cite{HerNJP12,KoePRL18} An interesting question with respect to particle stability is how many atoms $N$ a metal cluster has to have in order to bind a certain number of additional electrons. The critical size $N_c$ increases rapidly with the number of surplus electrons $z$. For example, tetra-, penta- and hexa-anionic gold clusters have been observed in Penning trap experiments.\cite{MarJPCC15} The corresponding $N_c$ indeed increases from 128 to 230 to 342, respectively. Further interest in polyanionic metal clusters concerns their electronic, optical and dynamical properties. In particular and in contrast to molecules, where the extra electron sticks to specific molecular sites, it is an open question, how the Coulomb barrier organizes in purely metallic nanoparticles and how the electronic shell structure develops. The method of choice to study the electronic structure of anions is photoelectron spectroscopy (PES) using ultraviolet laser radiation.\cite{HoJCP90,GanJCSFT90,TayJCP92,ThoPRL02,IssARPC05} So far, the magnetic bottle electron time-of-flight\cite{WanRSI99,MatJACS08} and the velocity map imaging method\cite{ChaPCCP14} have been applied to resolve the electronic properties of polyanions as well as their ultrashort dynamics.\cite{HorPRL12}  

Several techniques have been developed to produce polyanionic systems, i.e., laser desorption,\cite{LimJACS91,StrIJMS01} sputtering,\cite{SchPRL90c} electron capture,\cite{ComPRL97} electron transfer,\cite{BolPRL98,LiuPRL04} collision-induced dissociation\cite{MaaIJMSIP89} and electrospray.\cite{LauJCSCC95} Methods like electrospray rely on the formation of higher negatively charged molecules at the source exit. Although the expansion of a liquid in the presence of a strong electric field has become a standard method,\cite{WanJCP15} the limited flexibility in the adjustment of the source conditions restricts the production of anions in desired negative charge states. Electron post-feeding in collisions, in contrast, suffers from low cross sections.\cite{LeiIJMSIP86,ComPRL97} An improvement of the latter concept arises from conducting the electron attachment to a precursor anion within a restricted volume for a longer time span, i.e. using ion traps.

In Penning traps, polyanions are formed by exposing the clusters to an electron bath, i.e., by attachment of simultaneously stored electrons to precursor anions.\cite{HerPS99} As a disadvantage, the range of the accessible cluster sizes of the singly charged precursors are restricted due to the "critical mass" limits.\cite{SchIJMSIP95a} These constraints do not apply to radio frequency quadrupole ion traps. However, simultaneous trapping of electrons is not feasible in harmonic rf fields. Only recently, electron attachment to anions in a linear Paul trap has been demonstrated by applying a digital 3-state trapping scheme,\cite{BanIJMS13b} whereby an electron beam is guided through the trap in potential-free time periods.\cite{MarHI15}

\begin{figure*}[tb]
	\resizebox{1.0\textwidth}{!}{\includegraphics{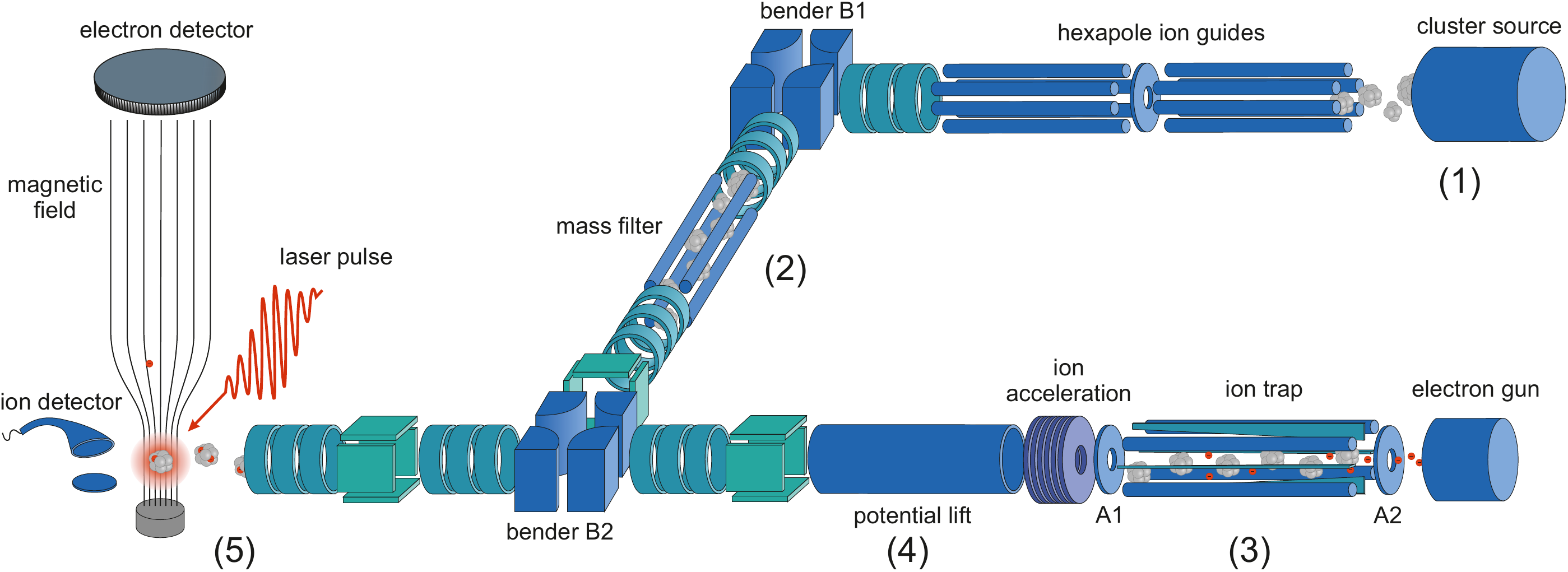}}
	\caption{Experimental setup for photodetachment experiments on size-selected cluster polyanions. In the present configuration metal cluster anions produced in a gas aggregation source of the Haberland type (1) are guided by hexapole rf fields towards an electrostatic quadrupole bender (B1), where anions are separated from neutrals and cations. After digital rf-based mass filtering (2), size-selected cluster anions enter a 3-state driven, linear Paul trap for polyanion production (3). Acceleration towards the photoelectron diagnostics is realized by applying the potential-lift technique (4). Tunable laser pulses serve to conduct photoemission experiments (5). The energy of the photoelectrons are determined by use of a magnetic bottle time-of-flight spectrometer. Various ion-optics elements guide the particles through the apparatus. See text for further details.}
    \label{fig:RasArXive22-f1}
\end{figure*}

In this contribution, we report about a setup, that combines  multiple electron attachment in a digital linear rf ion trap to size-selected particles with photoelectron spectroscopy. By the example of silver clusters, the capabilities with respect to target preparation and electron diagnostics is demonstrated. The paper is organized as follows. Sec.~\ref{sec:II-A} gives an overview of the entire setup. In Sec.~\ref{sec:II-B} the possibility to extent the available cluster size range by digital mass filtering is discussed. Sec.~\ref{sec:II-C} illustrates the method of in-trap electron attachment in a 3-state driven, linear Paul trap. In the Sec.~\ref{sec:II-D} the extraction and the ability to provide a high target density in the interaction region is described. Finally in Sec.~\ref{sec:II-E}, a procedure is described which allows for a comprehensive energy calibration of the time-of-flight electron spectra by taking advantage of tunable laser radiation. To demonstrate the capabilities of the setup, Sec.~\ref{sec:III} gives selected examples for studies on single (Sec.~\ref{sec:III-A}) and sequential multiphoton (Sec.~\ref{sec:III-B}) electron emission to determine  electronic cluster properties. Further, in Sec.~\ref{sec:III-C}, photoelectron spectra are analyzed with respect to the optical response.

\section{Experimental Setup}
\label{sec:II}

\subsection{Overview}
\label{sec:II-A}

The entire setup for photoelectron spectroscopy on size and charge-state selected polyanionic clusters is outlined in Fig.~\ref{fig:RasArXive22-f1}. Metal particles are produced in a magnetron sputtering gas aggregation source (1).\cite{HabJVSTA92} The molecular beam comprises a broad size distribution of neutral ($M_N$), positively  ($M_N^+$) as well as negatively ($M_N^-$) charged clusters. Radio frequency driven hexapole ion guides transfer the clusters towards an electrostatic bender (B1) (Extrel, Large Quad Deflector),\cite{ZemRSI77,FarRSI85} which separates the negative from the neutral and positive clusters and serves as a rough energy filter. The deflected anions enter a quadrupole mass filter with an energy of about $\SI{25}{\eV}$ (2). Typical currents after size selection are in the order of $\SI{10}{\pA}$. These clusters are guided into a linear, 3-state digital Paul trap (3).\cite{BanIJMS13a} Field-free time slots in the rf-cycle allow low energy electrons to pass the trap unhindered, whereby a weak superimposed magnetic field (not shown in Fig.~\ref{fig:RasArXive22-f1}) serves to guide the electrons  along the trap axis. Inelastic collisions of the electrons with stored ions and buffer gas lead to (multiple) electron attachment to the clusters.

After pulsed extraction, the potential-lift technique (4)\cite{MarEPJD11} is applied to accelerate the anions towards the photoelectron diagnostics (5). The charge-state dependent acceleration separates the size-selected polyanionic clusters in time with respect to $z$. In the interaction region the polyanions are exposed to laser pulses from a tunable \SI{1}{\kHz} nanosecond (\SIrange[range-phrase= -]{3}{6}{\ns}) laser system (EKSPLA, model NT242-SH/SFG), triggering photoemission. Typical photon energies applied in the experiments range from $E_{\text{ph}}=\SIrange[range-phrase=\text{ to }]{2.00}{5.82}{\eV}$. A magnetic bottle time-of-flight (TOF) electron spectrometer serves to determine the photoelectron energies providing information about the electronic and optical properties. Parts of the present setup are standard modules being used in cluster physics experiments, e.g., cluster source\cite{HabJVSTA92} and electron spectrometer.\cite{KruJPE83} In the following sections we thus focus on issues, which are crucial to our current experiment.

\begin{figure*}[tb]
	\resizebox{1.0\textwidth}{!}{\includegraphics{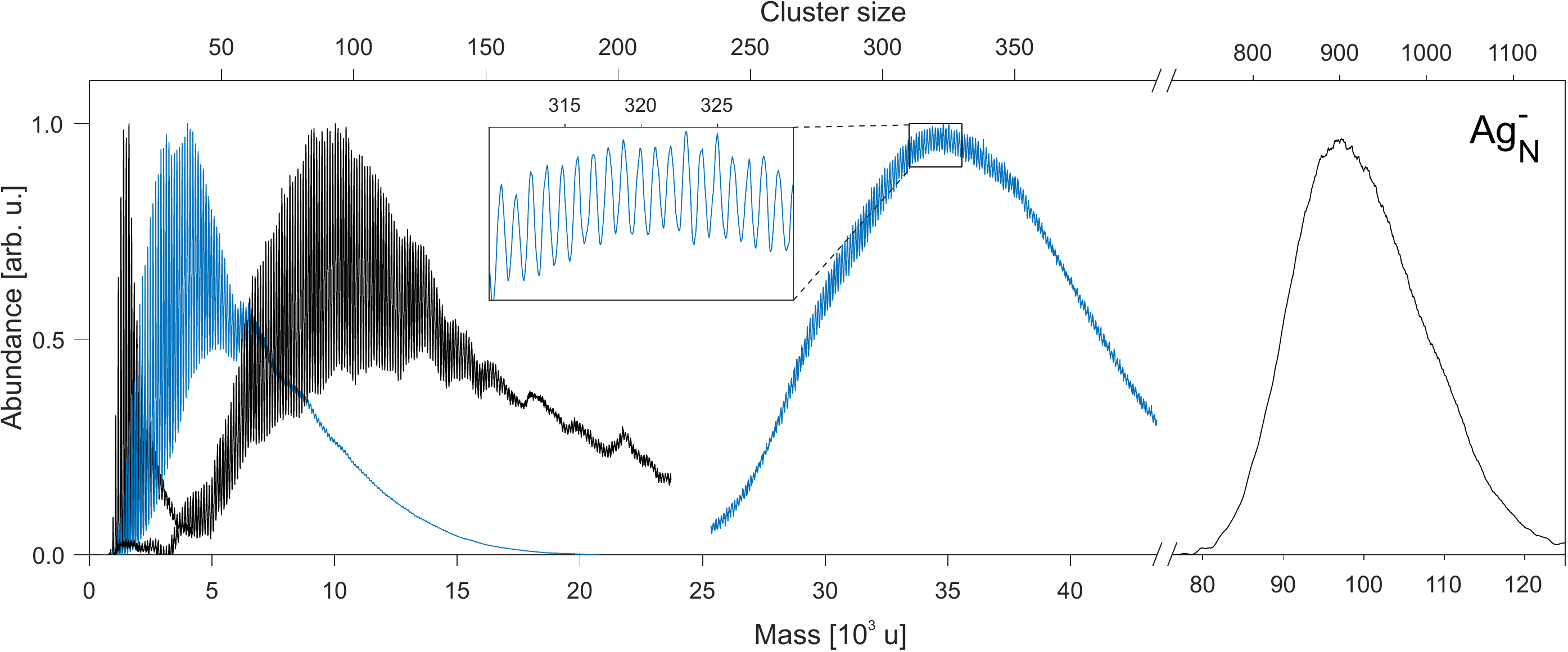}}
	\caption{Selected examples of Ag$_N^-$ mass spectra, demonstrating the capabilities of the mass filter driven by a rectangular wavefrom to provide a broad range of cluster sizes. The inset demonstrates the resolving power of the digital mass filtering. For better comparison, each spectrum have been normalized.}
    \label{fig:RasArXive22-f2}
\end{figure*} 
 
\subsection{Cluster size selection}
\label{sec:II-B}

The molecular beam apparatus is equipped with a commercial quadrupole mass filter for size-selection (Extrel, model GP-203, 9.5~mm Tri), originally driven by a standard harmonic high frequency generator (Extrel, model 150-QC, $\nu_{\text{rf}}=\SI{440}{\kHz}$). The instrument allows for mass filtering up to $\SI{16000}{\amu}$, corresponding to a silver cluster size of $N=148$ atoms for the example of silver. Since the attachment of surplus electrons depends on the cluster size,\cite{SchS95, YanPRL01} an extension of the mass range is essential to investigate the properties of higher negative charge states. As outlined in Sec.~\ref{sec:III-B}, silver clusters exceeding 800 atoms are required to obtain a polyanionic charge state of $z=8$. To be able to select clusters in this size range and beyond, the harmonic radiofrequency has been replaced by a rectangular waveform. This waveform is provided by two high voltage switches (CGC Instruments, model 19"–AMX1500–3E) that are supplied with a positive and a negative DC voltage $\pm U_{rf}$ (FuG, model MCP 350-2000). The switching frequency $\nu_{rf}$ can be adjusted using a waveform generator (Rigol, model DG1022Z).

The tunability of the system in terms of size range and resolution is restricted by the available setting parameters, which are currently limited to $U_{\text{rf}}\le\SI{750}{\V}$ and $\nu_{\text{rf}}\le\SI{250}{\kHz}$ balancing mass resolution and yield. In the present experiments, a mass resolution of $m/\Delta m\approx 320$ is achieved. With the digitally driven rf-fields, mass selection was applied to silver clusters as large as $N=1200$. Fig.~\ref{fig:RasArXive22-f2} shows examples of Ag$_N^-$ mass distributions, obtained for different source  settings. Hence, size-selected anionic clusters covering a broad size range are provided for the generation of Ag$_N^{z-}$. 

\subsection{Electron attachment to anionic clusters}
\label{sec:II-C}

\begin{figure}[b]
	\resizebox{1.0\columnwidth}{!}{\includegraphics{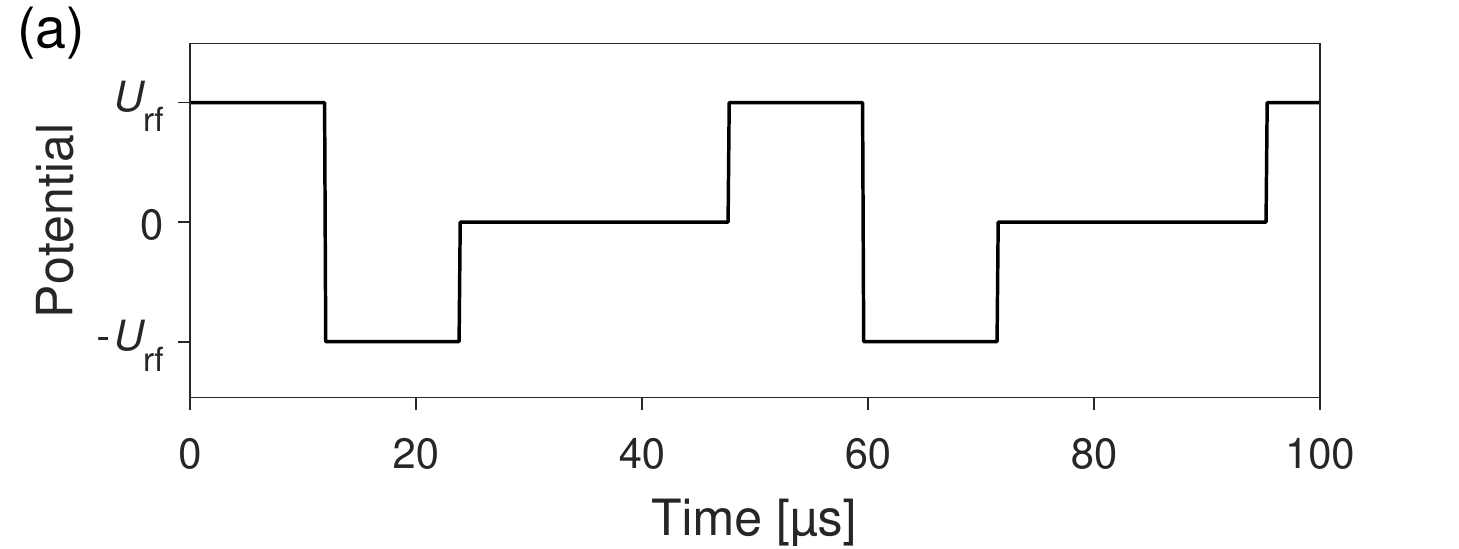}}
	\resizebox{1.0\columnwidth}{!}{\includegraphics{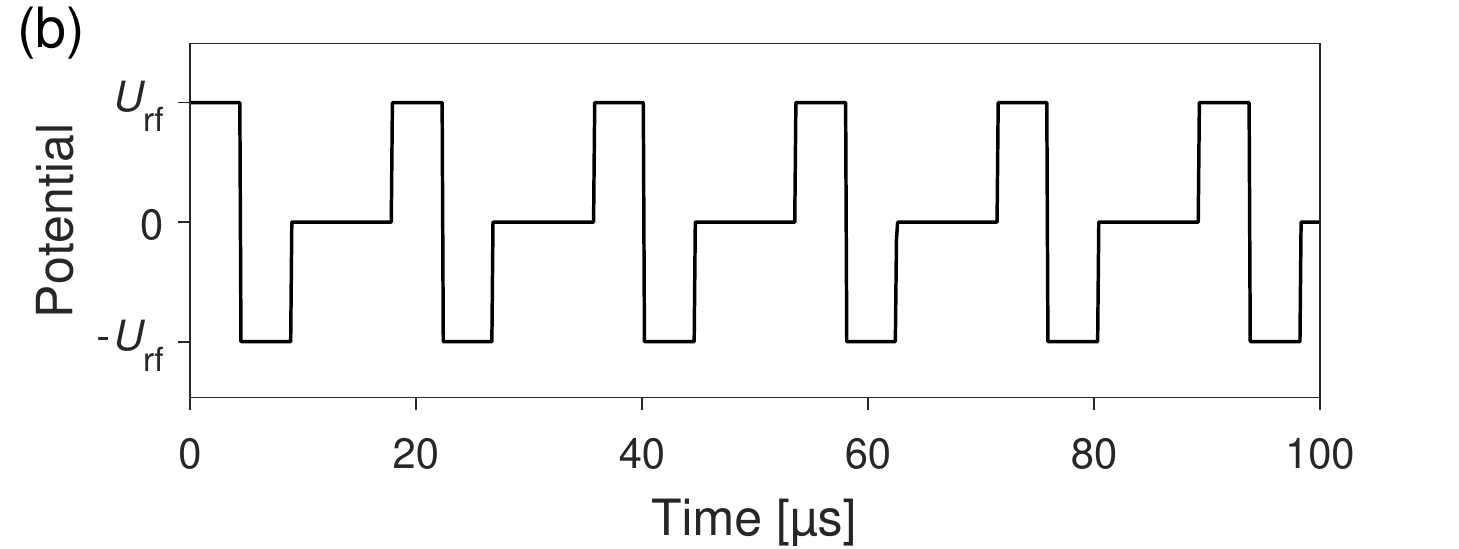}}
	\caption{Example for switching the radio frequency from cluster accumulation (a) to electron attachment mode (b). For  Ag$_{800}^{1-}$ the rf frequency is changed from $\nu_{\text{rf}}=\SI{21}{\kHz}$ to $\SI{56}{\kHz}$ in order to effectively run the attachment process eventually leading to the formation of Ag$_{800}^{7-}$.} 
    \label{fig:RasArXive22-f3}
\end{figure}

\begin{figure*}[t]
 	\resizebox{0,7\textwidth}{!}{\includegraphics{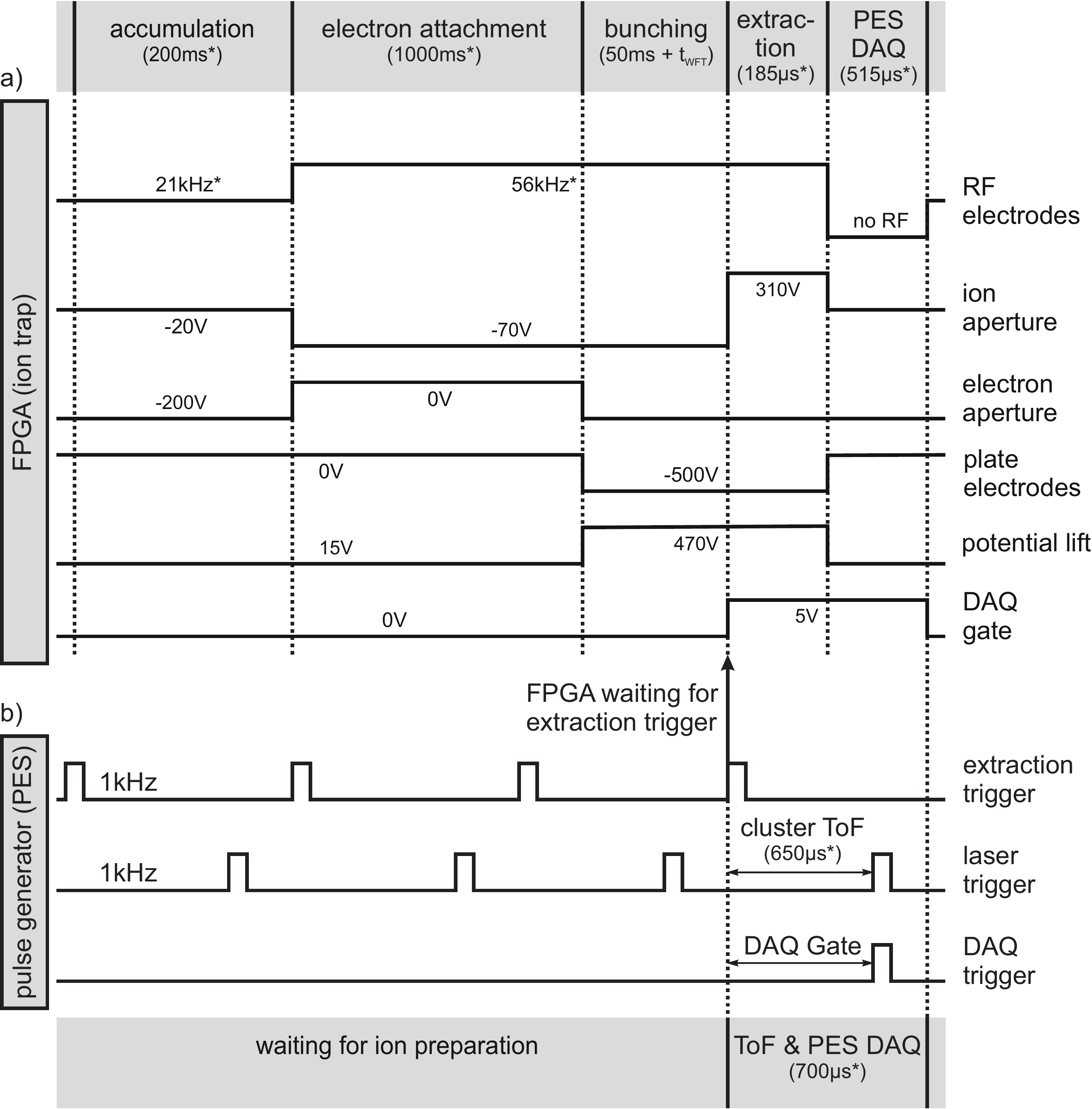}}
	\caption{Timing diagram of the experimental cycle to study photoemission from polyanionic metal clusters, which combines the potentials for the trap electrodes controlled by a field programmable gate array (FPGA) (a) and the trigger signals for synchronizing the PES provided by a pulse generator (Quantum Composer, model 9530) (b). The values labelled with (*) are size and/or charge state dependent and exemplarily given for Ag$_{800}^{7-}$. The dashed lines seperate the different sections, which are used within an experimental cycle. The two timing patterns are synchronized by the extraction trigger provided by the pulse generator (PES). After accumulation and electron attachment, the FPGA waits ($t_{\text{WFT}}\leq 1$~ms) in the bunching period after a fixed time of 50ms for the next pulse of the $\SI{1}{\kHz}$ extraction trigger signal.}
    \label{fig:RasArXive22-f4}
\end{figure*}

After mass selection, singly negatively charged clusters are accumulated in a home-build, linear Paul trap (Fig.~\ref{fig:RasArXive22-f1}). The trap consists of four circular rod electrodes with a length of $\SI{250}{\mm}$, a radius of $\SI{9}{\mm}$  and a minimal distance of $\SI{8}{\mm}$ to the trap axis. The radial confinement is provided by digital rf potentials on the rods ($U_{\text{rf}}=\SI{270}{\V}$) and using frequencies of up to $\nu_{\text{rf}}=240$\,kHz, depending on cluster size. The ion trap operates in a 3-state digital radio frequency mode\cite{BanIJMS13a,BanIJMS13b} where the rod potentials are switched between $\pm U_\text{rf}$ in one half of a period and during the other half all rods are on ground potential, see Fig.~\ref{fig:RasArXive22-f3}. The individual levels are provided by high voltage power supplies (FuG, model MCL 140-1250). The corresponding timings are adjusted by a waveform generator (Rhode \& Schwarz, model AM300) driving 3-state high voltage switches (CGC Instruments, model AMX1500-3F). The axial confinement is achieved by dc-potentials applied to electrodes located at both ends of the trap (ion aperture: A1, electron aperture: A2, see Fig.~\ref{fig:RasArXive22-f1}). 

The timing of the experimental cycle is depicted in Fig.~\ref{fig:RasArXive22-f4}. Size-selected clusters are accumulated in the trap by argon buffer gas cooling at a pressure of $\SI{e-3}{\milli\bar}$. Floating all trap components improves the accumulation as it leads to a lower excess energy and thus to a faster cooling of the clusters. Thereafter, in-trap multiple electron attachment\cite{MarHI15} is used to produce Ag$_N^{z-}$ from the initially accumulated anions. By switching A2 to ground potential, electrons from a thermionic emission gun enter the trap volume during the field-free time slots of the trapping period (Fig.~\ref{fig:RasArXive22-f4}). A weak superimposed magnetic field ($\SI{10}{\milli\tesla}$) guides the electrons along the trap axis, while the ion storage conditions are not affected significantly. Optimal attachment conditions were found at electron energies of 20 eV.

In contrast to electron capture by neutral atoms under single collision conditions, i.e. the Langevin formalism,\cite{RabJPCA11} the sequential electron attachment under complex trap conditions has not yet been modelled. In addition, each electron capture ($\text{Ag}_N^{z-} + e^-\rightarrow\text{Ag}_N^{(z+1)-}$) is accompanied by an increase of the Coulomb barrier of the polyanion. Hence, the most probable electron energy allowing for successful attachment is $z$-dependent. Thermalization of the energetic electrons motion by collisions with the buffer gas or electron impact ionization of argon producing low energy secondary electrons are processes probably leading to electron capture. Most likely, attachment is most effective for electrons having energies just above the Coulomb barrier.

The subsequent attachment of further excess electrons changes the mass-to-charge ratio of the clusters. Therefore, the storage conditions of the trap have to be readjusted to ensure stable confinement of the final polyanionic charge state $z_{\text{f}}$. This is achieved by increasing the radio frequency for monoanion accumulation by a factor of $\sqrt{z_{\text{f}}}$.\cite{GerACP92} For example, the experimentally optimized storage value of $\nu_{\text{rf}}=\SI{21}{\kHz}$ for Ag$_{800}^{1-}$ is tuned to $\SI{56}{\kHz}$ to trap Ag$_{800}^{7-}$, see Fig.~\ref{fig:RasArXive22-f3}. It was found that readjusting $\nu_{\text{rf}}$ has no severe effect on the further storage efficiency of the thermalized monoanions. Note, that the attachment process results in a distribution of charges, see Sec.~\ref{sec:II-D}.

The procedure requires about \SIrange[range-phrase= -]{800}{1000}{\ms} in order to maximize the yield of Ag$_{800}^{7-}$, whereas only $\SI{50}{\ms}$ are sufficient to obtain a strong signal from Ag$_{800}$ dianions. The attachment time optimized for a given charge state determines the duration of the experimental cycle. Interestingly, when optimizing for a maximum yield of heptamers, still more hexamer ions are detected, see Fig.~\ref{fig:RasArXive22-f5}. This possibly indicates a finite lifetime on the order of the experimental cycle duration due to the metastable nature of Ag$_{800}^{7-}$.\cite{IweJCP21} Note that due to multiple collisions and the storage conditions, this experimental lifetime only provides a lower limit with respect to the tunneling lifetime of undisturbed clusters of low internal excitation energies.

\subsection{Polyanion extraction}
\label{sec:II-D}

After the electron attachment, the polyanions are bunched near the ion aperture (A1, Fig.~\ref{fig:RasArXive22-f1}) by applying a negative potential to the plate electrodes\cite{LobEJMS00} mounted between the rf rods (Fig.~\ref{fig:RasArXive22-f1} and Fig.~\ref{fig:RasArXive22-f4}). Subsequently, the ion aperture is switched to a positive potential of $\SI{310}{\V}$ (Fig.~\ref{fig:RasArXive22-f4}) which leads to a pulsed extraction into the ion acceleration unit, which includes a potential lift tube (Fig.~\ref{fig:RasArXive22-f1}). The polyanions enter the tube at typically $\SI{470}{\V}$. While the ions are inside, the potential is switched back to $\SI{15}{\V}$ (Fig.~\ref{fig:RasArXive22-f4}). Hence, the ions keep their kinetic energies, when they exit the tube. The delay $\Delta t_{\text{lift}}$ between switching the ion aperture and the potential lift must be adjusted in such a way, that the cluster bunch locates mainly within the tube.\cite{MarEPJD11}. The optimal $\Delta t_{\text{lift}}$ depends on cluster size and charge state under investigation, e.g. 500~$\mu$s for $\text{Ag}_{800}^{1-}$. Additionally, the ion bender B2 (Fig.~\ref{fig:RasArXive22-f1}) is switched to ground potential. The polyanions can thus pass unhindered towards the diagnostics, i.e., ion detection and photoelectron spectrometer. 

\begin{figure}[tb]
	\resizebox{1.0\columnwidth}{!}{\includegraphics{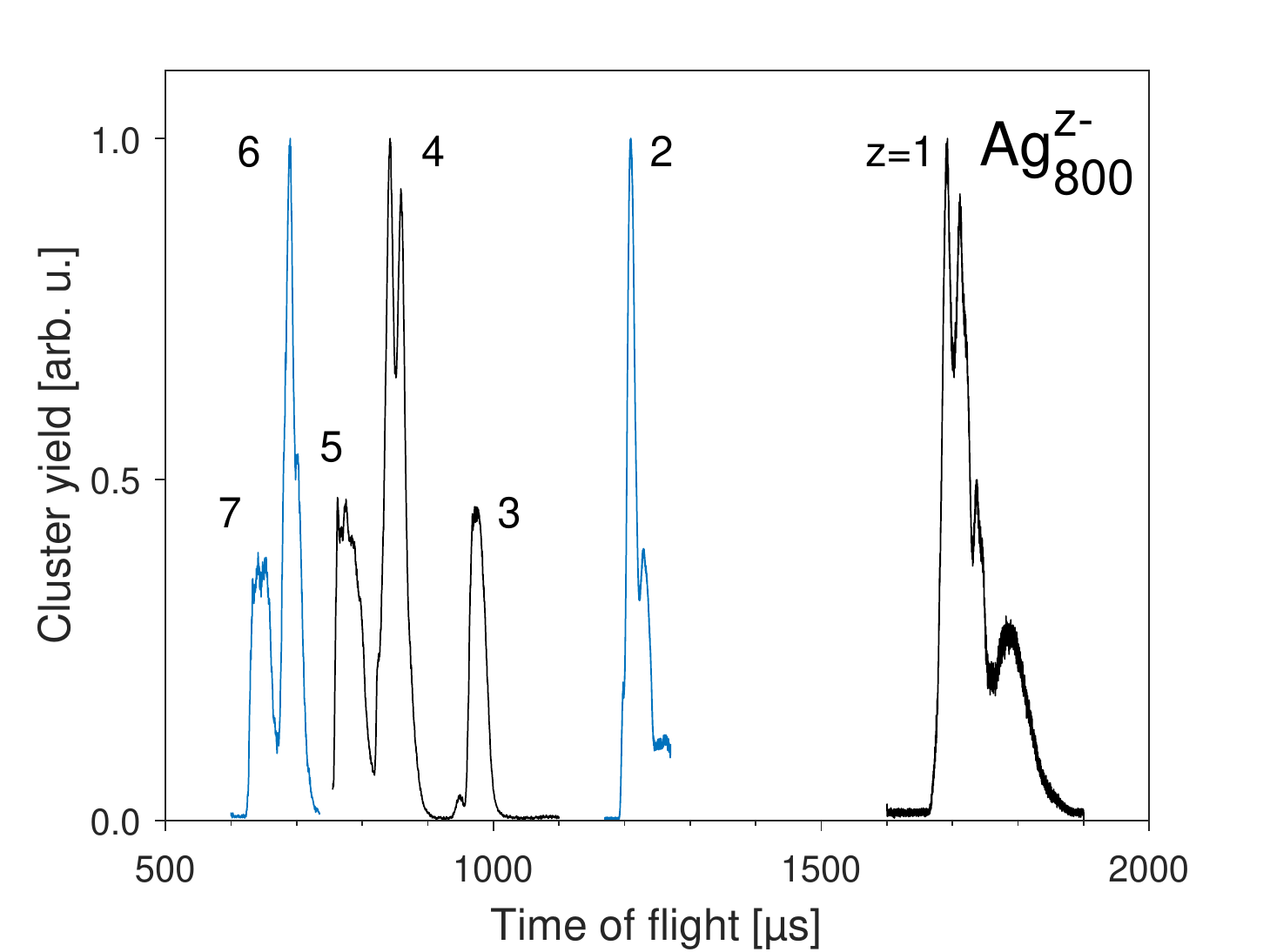}}
	\caption{Ion detector signals (see Fig.~\ref{fig:RasArXive22-f1}) of charge-state distributions obtained under different storage, electron attachment and polyanion extraction conditions. By adjusting parameters as $\nu_{rf}$, duration of electron attachment and $\Delta t_{\text{lift}}$, the yields of selected charge states are optimized, i.e. for $z=$ 1, 2, 4 and 7 in the selected spectra. The spectra have been normalized to their maximum.} Different colors are used to distinguish between neighbouring spectra.
    \label{fig:RasArXive22-f5}
\end{figure}

The attachment process in the trap results in different charge states of the mass-selected clusters. For the example of Ag$_{800}^{z-}$, Fig.~\ref{fig:RasArXive22-f5} shows typical time-of-flight spectra recorded with the channeltron ion detector behind the interaction region, see Fig.~\ref{fig:RasArXive22-f1}. Due to the charge-state dependent ion acceleration, the different polyanions separate from each other because of their time-of-flight to the detector and the different charge states are resolved. Note that the ion extraction is efficient only for a narrow $z$-range (given by $\Delta t_{\text{lift}}$) and therefore do not reflect the actual charge-state distribution in the trap.\cite{MarEPJD11} For this reason, the trap and extraction parameters have been adjusted for the individual spectra. 

For PES on mass and charge-state selected clusters, the delay between extraction and laser trigger is set according to the time-of-flight of the species of interest, see Fig.~\ref{fig:RasArXive22-f4} (b). Whereas extraction trigger and laser system operate at $\SI{1}{\kHz}$, the target preparation is much slower ($\approx$1-20~Hz). Therefore, only the extraction trigger following the bunching period is actually used (Fig.~\ref{fig:RasArXive22-f4}). To efficiently record photoemission spectra (Sec.~\ref{sec:II-E}), a sufficiently high and almost constant shot-to-shot target density in the interaction region is beneficial. First of all, the accumulation of cluster anions in the Paul trap effectively reduces the initial signal fluctuations originating from the source. This improvement is maximized when the trap is filled up to the Coulomb limit. In addition, for the  photoemission experiments the maximum yield of the extracted ion pulse has to match with the arrival time of the laser pulse. It was found that keeping the rf phase at the time of extraction constant, optimizes the ion pulse train.

\subsection{Photoelectron diagnostics}
\label{sec:II-E}

A magnetic bottle time-of-flight electron spectrometer,\cite{SenPRL09} see Fig.~\ref{fig:RasArXive22-f1}, is used to record photoelectron signals from cluster polyanions. The total length of the time-of-flight region is $\SI{1.2}{\m}$. The magnetic field configuration collects electrons produced in the interaction region over nearly the full solid angle. The magnetic bottle includes a permanent magnet ($\SI{5}{\mm}$ diameter) with an estimated field strength of $\SI{1}{\tesla}$ and a 1\,mT guiding field inside a $\SI{100}{\mm}$ diameter drift tube. To minimize external magnetic stray fields, the electron drift tube is surrounded by a $\mu$-metal shield. In order to improve the detection efficiency at kinetic energies below $\SI{0.2}{\eV}$, a weak electrostatic potential at the permanent magnet\cite{MatRSI11} can be used to accelerate the electrons towards a 40\,mm diameter, 3-stack multi-channel plate detector.

Recording photoelectron spectra by time-of-flight techniques in particular on cluster anions are often hampered by a low number of reference values available to calibrate the energy scale. Using the broad wavelength tunability of our laser system, however, spectra of various atomic gases and selected cluster anions can serve as database for calibration. Resonance enhanced 2-photon ionization-induced electron emission (R2PI) is applied to divalent neutral atoms like Mg or Ca,\cite{NIST} being emitted from a resistively heated oven. Due to the thermal evaporation, the entire interaction region is filled with the metal vapor. Although the vapor density in the interaction region is low, the strong atomic absorption at  resonance\cite{NIST} yields sufficiently high electron signals. As the measurements on the effusive atomic beam are conducted under static conditions, the data acquisition is solely limited by the laser pulse repetition rate of $\SI{1}{\kHz}$, lowering considerably the time for recording the R2PI spectra. Typically $10^5$ laser shots at a rate up to 10 electrons per shot are sufficient to obtain reliable data for the calibration procedure.

\begin{figure}[tb]
	\resizebox{1.0\columnwidth}{!}{\includegraphics{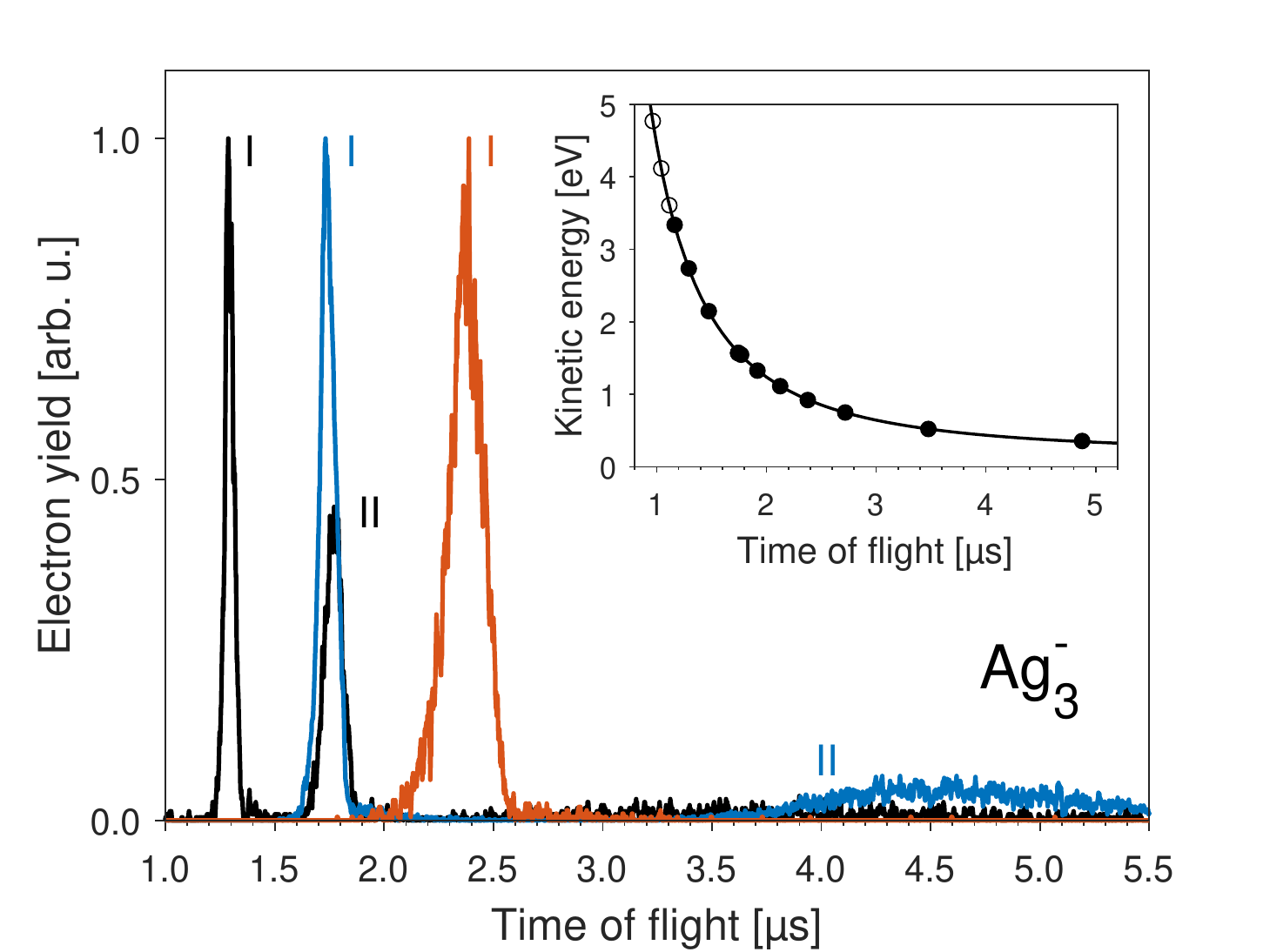}}
	\caption{Examples of photoelectron TOF spectra obtained from Ag$_3^-$ recorded at selected photon energies of $E_{\text{ph}}=\SI{5.17}{\eV}$ (black), $\SI{4.00}{\eV}$ (blue) and $\SI{3.35}{\eV}$ (red), which are used to calibrate the kinetic energy axis. The spectra show signals from electronic levels that correspond to binding energies of $\SI{2.43}{\eV}$ (I) and $\SI{3.62}{\eV}$ (II).\cite{HanJCP95} The resulting calibration curve ($E_{kin}=a/t^2+c$, black line) extracted from the spectra of non-accelerated ($\bullet$) and accelerated ($\circ$) electrons is shown in the inset, see text for details. The spectra have been normalized for better comparision.} 
   \label{fig:RasArXive22-f6}
\end{figure}

Regularly photoemission from Ag$_3^-$ is used to check the calibration. The corresponding electron spectrum exhibits narrow and well-resolved peaks at binding energies of $E_{\text{bin}}=2.43$ and $\SI{3.62}{\eV}$,\cite{HoJCP90,HanJCP95} see Fig.~\ref{fig:RasArXive22-f6}. Limited by the maximum available photon energy of $E_{\text{ph}}^{\max}=\SI{5.82}{\eV}$, the procedure allows to assign electron kinetic energies up to $E_{\text{kin}}=E_{\text{ph}}^{\max}-E_{\text{bin}}=\SI{3.39}{\eV}$.
In order to extend the calibration range to higher energies of about $\SI{5}{\eV}$, electrons are accelerated by applying a negative voltage at the permanent magnet.\cite{MatRSI11} By analyzing  electron time-of-flight spectra taken at various laser photon energies a calibration curve is determined, see inset in Fig.~\ref{fig:RasArXive22-f6}. 

\section{Results and Discussion}
\label{sec:III}

\subsection{Photoemission from \textrm{Ag}$_{800}^{4-}$}
\label{sec:III-A}

\begin{figure}[tb]
	\resizebox{1.0\columnwidth}{!}{\includegraphics{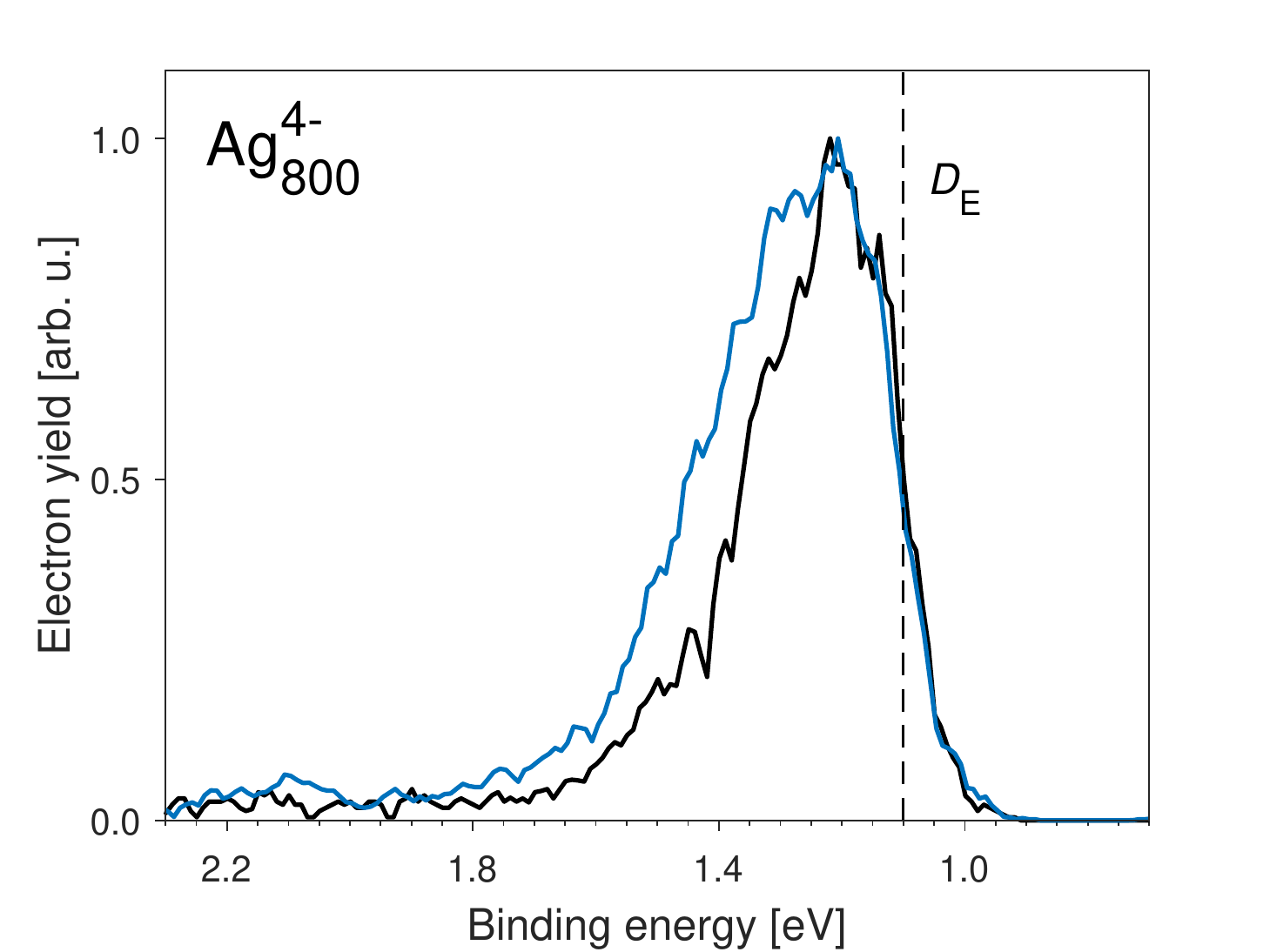}}
	\caption{Photoelectron spectra of Ag$_{800}^{4-}$ obtained at photon energies of $E_{\text{ph}}=\SI{3.18}{\eV}$ (black) and $\SI{3.31}{\eV}$ (blue). The dashed line denotes the detachment energy $D_\text{E}=\SI{1.10(5)}{\eV}$ of Ag$_{800}^{4-}$.\cite{IweJCP21} The high binding energy wing provides information of the Coulomb barrier.\cite{MarPRL21} The spectra have been normalized to their maximum.}
    \label{fig:RasArXive22-f7}
\end{figure}

In the interaction region of the electron spectrometer, the size and charge-selected polyanions are exposed to pulsed laser radiation. Fig.~\ref{fig:RasArXive22-f7} (black) shows the photoelectron spectrum of Ag$_{800}^{4-}$ taken at a photon energy of $E_{\text{ph}}=\SI{3.18}{\eV}$. The signal stems from electrons in energy levels covering a range of about $\SI{0.8}{\eV}$. As outlined in,\cite{IweJCP21} the corresponding detachment energy is determined from the steepest slope of the signals at low binding energies, giving $D_\text{E}=\SI{1.10(5)}{\eV}$, which is attributed to the highest occupied cluster orbital. Signals of the electronic structure up to binding energies of $E_\text{bin}=\SI{3.18}{\eV}$ are expected. But the signal already decreases above $E_\text{bin}=\SI{1.2}{\eV}$ and fades out at around $E_\text{bin}=\SI{1.7}{\eV}$. The falling edge at higher binding energies originates from the presence of a Coulomb barrier, which prevents direct emission of low energy electrons.\cite{WanPRL98a, MarPRL21} However, by means of tunneling through the barrier, electrons with lower energy exit the cluster and contribute to the spectrum. The impact of the Coulomb barrier on the left-hand side of the spectrum is illustrated, when recording data at higher photon energies, e.g. 3.31\,eV, see Fig.~\ref{fig:RasArXive22-f7} (blue). The left-hand wing shifts according to the difference of the photon energies and confirms the presence of a barrier which suppresses the emission of low energy electrons.\cite{MarPRL21} The yields of tunneling electrons, however, depend on height and shape of the barrier potential as well as the level density. In addition, one has to take into account a possible energy dependence of the photoabsorption cross-section.

\begin{figure}[tb]
	\resizebox{1.0\columnwidth}{!}{\includegraphics{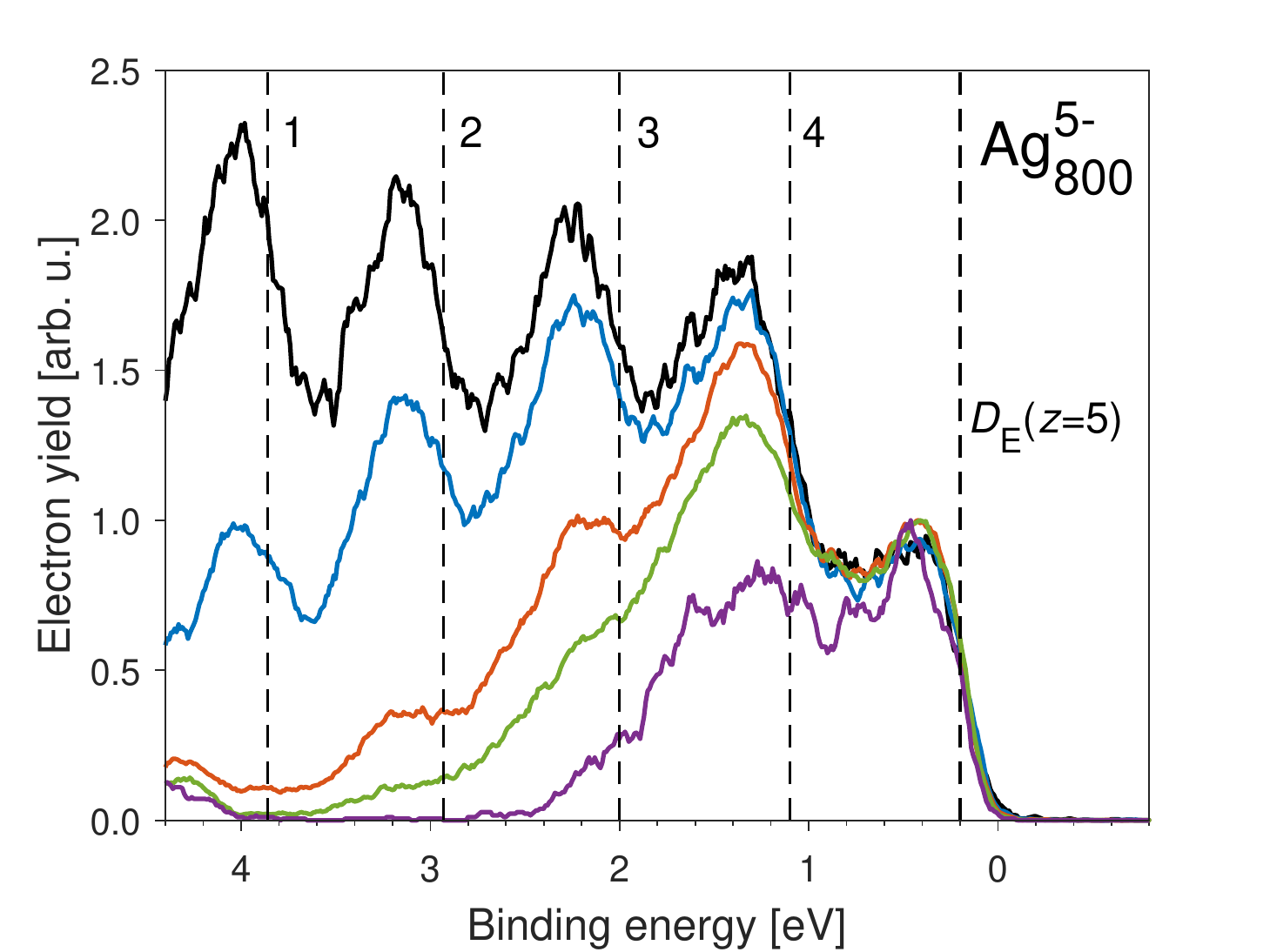}}
	\caption{Photoelectron spectra of Ag$_{800}^{5-}$ obtained under multiphoton absorption conditions ($E_\text{ph}$=\SI{4.66}{\eV}). For the measurements, laser pulse energies of $\SI{200}{\uJ}$, $\SI{90}{\uJ}$, $\SI{30}{\uJ}$, $\SI{15}{\uJ}$ and $\SI{0.5}{\uJ}$ (top to bottom) are applied. With increasing laser intensity, signals from sequential photoemission contribute and lead to distinct "Coulomb staircase" features.\cite{AstPRB02} The dashed lines denote the detachment energies of the different charge states $z=\SIrange[range-phrase= -]{1}{5}{}$.\cite{IweJCP21}}
    \label{fig:RasArXive22-f8}
\end{figure}

\subsection{Multiphoton processes}
\label{sec:III-B}

In the following, the impact of the laser intensity on the photoelectron spectra is analyzed for the example of Ag$_{800}^{5-}$, see Fig.~\ref{fig:RasArXive22-f8}. When increasing the laser fluence, additional peaks show up in the spectrum, with the maxima separated by roughly the same energy of $\SI{0.9}{\eV}$ (Fig.~\ref{fig:RasArXive22-f8}). The intensity dependence reveals, that the individual maxima originate from sequential photoemission, i.e., 

\begin{equation}
  \nonumber
  \textrm{Ag}_{800}^{z-}+\hbar\omega\rightarrow\textrm{Ag}_{800}^{(z-1)-}+e^-~~~z=5,\ldots,1
\end{equation}

Each peak thus reflects the photoelectron spectrum of Ag$_{800}^{z-}$ emitted from different charge states. Obviously, the energy shifts refer to the cluster charging energy. For example, the peak at $E_\text{bin}=\SI{1.3}{\eV}$ stems from Ag$_{800}^{4-}$ photoemission. Supporting evidence for this assignment is obtained, when comparing the signal to the tetraanion spectrum shown in Fig.~\ref{fig:RasArXive22-f7}.  To obtain more specific values, detachment energies have to be determined in single-photon experiments on each $z$ separately to avoid interferences with other charge states.\cite{IweJCP21} Further, between the successive photon absorption events fragmentation has to be considered as it may compete to detachment.\cite{ZheJCP86} Despite the rough approximation in determining the charging energies, the good agreement of the charging energy with the values obtained in\cite{IweJCP21} suggests that each absorption of a photon triggers the emission of only a single electron. 

\subsection{Optical spectra}
\label{sec:III-C}

\begin{figure}[t]
	\resizebox{1.0\columnwidth}{!}{\includegraphics{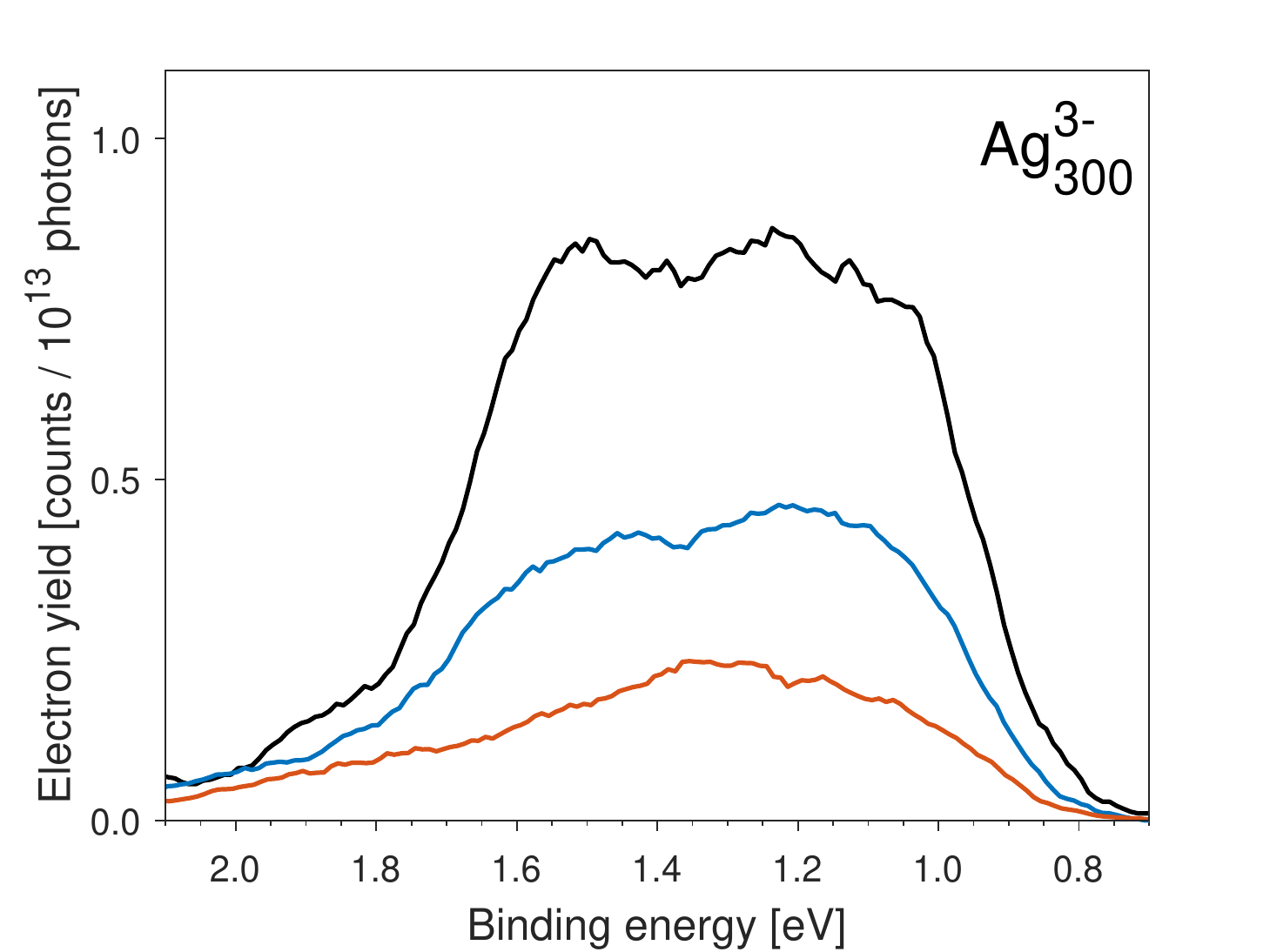}}
	\caption{Photoelectron spectra of Ag$_{300}^{3-}$ obtained at photon energies of $E_{\text{ph}}=\SI{3.65}{\eV}$ (black), $\SI{3.76}{\eV}$ (blue) and $\SI{3.87}{\eV}$ (red). In order to highlight the impact of the photodetachment cross section, the yields have been normalized to the laser fluence.}
    \label{fig:RasArXive22-f9}
\end{figure}

Recording wavelength-dependent photoelectron spectra offers the possibility to collect further information about the optical response of small particles. Moreover polyanions allow to extend the studies towards anionic charge-state dependent effects. In order to demonstrate the feasibility, Fig.~\ref{fig:RasArXive22-f9} shows photoelectron yields of Ag$_{300}^{3-}$ for selected photon energies, where the spectra have been normalized with respect to the laser fluence. We note an increase in the overall yields when lowering the photon energy from $\SIrange[range-phrase=\text{ to }]{3.87}{3.65}{\eV}$. Under the reasonable assumption, that the emission of electrons is directly linked to photoabsorption, the overall yields represent a measure of the absorption cross sections. The development in the cross sections stems from the collective oscillation of the delocalized electrons in silver, i.e., the Mie plasmon resonance $\hbar\omega_{\text{Mie}}$,\cite{MieAP08,HeePRL87} which in the small particle dipole limit gives a value of $\hbar\omega_{\text{Mie}}=\SI{3.5}{\eV}$.\cite{CharCRT98} Hence, there is strong indication that the plasmon is probed. The impact of the polyanionic charge states on the optical response has not been studied yet. Evidence for a presumable $z$-dependence in higher negatively charged clusters arises from measurements on monocations and -anions.\cite{TigPRA93,TigCPL96} Currently, studies are in progress, to systematically investigate the optical response of Ag$_N^{z-}$.\cite{IwePCCP22}

\section{Conclusions}
In summary, a novel setup is introduced, which allows detailed studies of highly negatively charged nanoparticles. The formation of specific polyanions is divided into subsequent steps executed in dedicated experimental components, i.e. size-selection in a digital mass filter, accumulation and charging in a digital 3-state, linear Paul trap. This permits us to optimize the individual preparation steps and in particular the electron attachment process, and thus to prepare clusters in a wide mass and charge-state range. By using tunable laser pulses, wavelength-dependent photoelectron spectroscopy is realized, which enables the study for, e.g. Coulomb barriers,\cite{MarPRL21} detachment energies,\cite{IweJCP21} sequential multiphoton detachment, as well as the optical response. In contrast to Penning traps, the storage capabilities of rf ion traps give access to much larger clusters. Due to the close connection between $N$ and $z$, which reflects in specific appearance sizes\cite{BanIJMS22} higher $z$-stages can be investigated. Further, since the setup does not include a heavy superconducting magnet, an rf-based solution is more suitable for conducting experiments at large scale facilities like free-electron lasers\cite{RosPR19,FukAPX20} or upcoming bright coherent soft-xray lasers, e.g. ELI-ALPS.\cite{KueJPB17} The application of a sequential target preparation procedure, which results in an intense pulsed beam of polyanions, however, is not restricted to metal clusters but can be extended to a wide range of polyanionic systems, simply by exchanging the particle source providing the anions. This includes synthetic polymers, which are of interest in industrial and biomedical applications.\cite{YesADDR04}

\section*{Acknowledgements}
K.R. and M.M. contributed equally to this work. We thank S. Bandelow for helpful discussions  in the early phase of the experiment. S. Lochbrunner provided us with a tunable laser system. The Deutsche Forschungsgemeinschaft (SFB652, TI 210/\,10) is gratefully acknowledged for financial support. M. M. acknowledges support by the International Helmholtz Graduate School for Plasma Physics (HEPP).

\end{document}